\documentstyle[12pt]{article}
\begin{document}
\input psfig
\begin{titlepage}
\begin{center}
\today     \hfill    WM-97-115\\

\vskip .25in

{\large \bf Supersymmetric model \\
of quasi-degenerate neutrinos}\footnote{This work
was supported in part by the National Science Foundation under grant 
PHY-96-00415.}

\vskip 0.3in

Christopher D. Carone and Marc Sher

\vskip 0.1in

{\em Nuclear and Particle Theory Group \\
     Department of Physics \\
     College of William and Mary \\
     Williamsburg, VA 23187-8795}


        
\end{center}

\vskip .1in

\begin{abstract}
We present a supersymmetric model of fermion masses, based on a non-Abelian
family symmetry and the Froggatt-Nielsen mechanism, that can account for the
solar and atmospheric neutrino problems via quasi-degenerate neutrinos. 
The model predicts that the ratio of neutrino mass squared splittings $\Delta
m^2_{12} / \Delta m^2_{23}$ is of order $m_s^2/m_b^2$, and the angles
$\theta_{12} \sim m_d/m_s$ and  $\theta_{23} \sim 1$, which are of the
desired orders of magnitude.  We discuss the implications of the 
flavor structure of the neutrino sector on superparticle masses and 
mixing angles.
\end{abstract}

\end{titlepage}

\newpage
\renewcommand{\thepage}{\arabic{page}}
\setcounter{page}{1}
\section{Introduction} \label{sec:intro} \setcounter{equation}{0}

The most striking feature of the quark and charged lepton masses is their
hierarchical structure.   An entire industry has developed in an attempt
to understand this pattern, most recently in the context of supersymmetric
theories \cite{sft}.  In this letter we will construct a supersymmetric 
model of flavor with the physics of the neutrino sector specifically in 
mind.

Given the hierarchy of the charged fermion masses, it is natural 
to assume that the neutrino masses follow a similar hierarchical 
structure:  $m_{\nu_e}<<m_{\nu_\mu}<<m_{\nu_\tau}$.  However, we will
consider an alternative possibility, that the three generations of
neutrinos are nearly degenerate, with small splittings sufficient to
explain (i) the solar neutrino problem via  $\nu_e$-$\nu_\mu$ 
MSW oscillations, and (ii) the atmospheric neutrino problem via 
$\nu_\mu$-$\nu_\tau$ mixing \cite{nra}. While these effects can be 
explained in models with hierarchical neutrino masses as well, it is 
important to explore all the possibilities, especially in light of the 
evidence from the SuperKamiokande experiment that neither the solar nor 
atmospheric neutrino problems are about to disappear \cite{superk}.  A 
number of interesting theories with quasi-degenerate neutrinos can be found in 
the recent literature \cite{fukugita,petcov,caldwell,binetruy}, and given
the improvement in the experimental situation, it is worthwhile to give this 
and other scenarios further consideration.

Our model is based on a global non-Abelian family symmetry that 
restricts the quark, lepton, neutrino, and scalar superparticle mass matrices
simultaneously; we don't consider the lepton sector in isolation,
unlike some of the other models on the market.  We assume flavor-symmetry 
breaking originates in the fermion Yukawa couplings via a sequential 
breaking of the flavor symmetry group\footnote{The symmetry-breaking 
scale is taken high enough so that the associated goldstone boson decay 
constants are above the lower bounds from direct collider searches for 
familons.}.  Hence, we avoid any ansatz for the Yukawa textures that 
cannot be motivated from symmetry considerations. Interestingly, the 
elements of the Yukawa matrices in our model that control right-handed 
rotations in the quark sector largely determine the neutrino mixing 
angles.  This provides another constraint on the form of the Yukawa 
textures.  We show that the model can solve problems (i) and (ii) above via
the seesaw mechanism, yielding the appropriate neutrino mass ratios and 
mixing angles in terms of ratios of quark Yukawa couplings.  In addition, 
the model solves the supersymmetric flavor problem by yielding superparticle 
degeneracies in the flavor symmetric limit.  With the origin of flavor 
symmetry breaking specified, we consider the detailed form of the 
superparticle mass matrices and state the distinctive supersymmetric 
signatures of the model.  

\section{The Model}
In the limit of vanishing Yukawa couplings, the supersymmetric standard model
has a global $U(3)^5$ flavor symmetry, with a separate U(3) factor acting on
each of the chiral superfields $Q$, $U$, $D$, $L$, and $E$.  One approach to
constructing models of flavor is to impose a horizontal symmetry that is a
subgroup of U(3)$_Q \times $U(3)$_U \times$ U(3)$_D$ alone, where U(3)$_Q$ and
U(3)$_D$ are assumed to act on the lepton fields $L$ and $E$.  In this way,
the known similarity between the down quark and charged
lepton masses \cite{gj} can be understood, up to factors of order
unity.  This approach was taken in Refs.~\cite{chm1,chm2} and will also be
adopted here. While this choice leaves a twofold ambiguity in assigning the 
$L$ and $E$ transformation properties, only one assignment, in which $L$ and 
$E$ transform under U(3)$_D$ and U(3)$_Q$, respectively, will lead to the 
desired pattern of neutrino masses and mixing angles.  We must first decide 
on a subgroup of U(3)$^3$ to use in constructing a model, and specify the 
transformation properties of the right-handed neutrino fields, $\nu$.

Since our goal is to construct a model of quasi-degenerate neutrino masses, we
will require the mass matrix of the light neutrino mass eigenstates to be
proportional to the identity in the flavor-symmetric limit.  Since this mass
matrix arises via the see-saw mechanism,
\begin{equation}
M_{LL} \approx M_{LR} M_{RR}^{-1} M_{LR}^T \,\,\, ,
\end{equation}
we can achieve degeneracy if the Dirac and Majorana mass matrices, $M_{LR}$
and $M_{RR}$, are individually proportional to the identity.   If the lepton
doublet $L$ is in a three dimensional representation of some subgroup of
U(3)$_D$, then we can achieve the desired form of $M_{LR}$ if $\nu$ transforms
in the ${\bf \overline{3}}$ (assuming the Higgs fields are singlets).  
However,  we also require $M_{RR}$ to be invariant under the flavor 
symmetry and proportional to the identity, which implies that 
${\bf \overline{3}} \sim {\bf 3}$; thus we seek a subgroup of U(3)$_D$ that
has real three-dimensional representations.   The largest subgroup that is
appropriate is SO(3), which we will adopt henceforth.  As for the 
part of the flavor symmetry that lives in the factors 
U(3)$_Q \times$ U(3)$_U$, we will take the largest symmetry possible, 
U(3)$_Q \times$U(3)$_U$ itself.  Thus, we begin with the maximal flavor 
symmetry group appropriate for our purposes,
\begin{equation}
G_f = U(3)_Q \times U(3)_U \times SO(3)_D  \,\,\, ,
\label{eq:gffull}
\end{equation}
where the three generations of left- and right-handed neutrinos transform as
three dimensional vectors under SO(3)$_D$.

The top quark Yukawa coupling breaks $G_f$ strongly down to
\begin{equation}
G_f^{eff}  = U(2)_Q \times U(2)_U \times U(1)_{Q_3-U_3} \times SO(3)_D \,\,\,,
\label{eq:effgf}
\end{equation}
where the U(1) factor rotates $Q_3$ and $U_3$ by an opposite phase.  This 
is the approximate flavor symmetry relevant at low energies, with the light
fermion Yukawa couplings treated as small symmetry breaking parameters.  
Notice that the large top quark Yukawa coupling does not break the SO(3) 
symmetry, so we don't expect any large deviation from degeneracy in the
neutrino sector. To properly construct the low-energy effective theory, we
consider all operators invariant under $G_f^{eff}$, with the light fermion
Yukawa couplings included as small symmetry-breaking parameters. 
Alternatively, we could directly impose $G_f^{eff}$ as the high-energy 
flavor symmetry, in which case an order one top quark Yukawa coupling 
would follow as a prediction of the theory. 

If we assume that the light fermion Yukawa couplings are the only source of
flavor symmetry breaking, we can estimate the symmetry breaking effect in any
operator of interest in the low-energy effective theory \cite{hr}.  Suppose
that the Yukawa couplings in the quark sector originate from fields
that transform simultaneously under pairs of the group factors 
in Eq.~(\ref{eq:gffull})
\begin{equation}
\Phi_u \sim ({\bf \overline{3}},{\bf \overline{3}}, {\bf 1})  \,\,\,\,\,\,\, ,
\Phi_d \sim ({\bf \overline{3}},{\bf 1}, {\bf 3})   \,\,\, ,
\label{eq:tprops}
\end{equation}
Here we state $G_f$ transformation properties of the fields for notational 
convenience only.  The reader should keep in mind that this is simply a 
shorthand for representing the set of fields contained in the $G_f^{eff}$ 
decomposition of the $\Phi$.  Using these fields, we may write down the 
following higher-dimensional superpotential interactions  
\begin{equation}
W = \frac{1}{M_F} (Q \Phi_u H_u U + Q \Phi_d H_d D) \,\,\, ,
\label{eq:supyuk}
\end{equation}    
where $M_F$ is the flavor scale.  We will specify the origin of these
operators below.  Yukawa couplings arise if the $\Phi$ acquire an 
appropriate pattern of vacuum expectation values (vevs).   For example, 
let us consider the following pattern of vevs, given in units of $M_F$: 
\begin{equation}
\Phi_d = 
\left(\begin{array}{ccc}
h_d  &   h_s\lambda &  h_b  V_{ub}  \\
h_d  &   h_s        &  h_b  V_{cb}  \\
h_d  &   h_b        &  h_b \\
\end{array}\right) 
\label{eq:texture1}
\end{equation}  
\begin{equation}
\Phi_u = 
 \left(\begin{array}{ccc}
h_u   &  h_s \lambda   &  h_s \lambda    \\
h_u   &  h_c           &  h_s            \\
h_u   &  h_s           &  h_t            \\  
\end{array}\right) \,\,\, .
\label{eq:texture2}
\end{equation}
Here, the $h_i$ are the quark Yukawa couplings, $\lambda \approx 0.22$ is the
Cabibbo angle, and the $V_{ij}$ are Cabibbo-Kobayashi-Maskawa (CKM) mixing 
angles.  Order one coefficients have been suppressed.  The diagonal elements 
yield the quark mass eigenvalues, while the entries that determine rotations 
on the left-handed down quark fields reproduce the CKM angles \cite{weird}.  
It is interesting that the remaining elements, which control right-handed 
rotations in the quark sector, will give us the desired flavor structure in 
the neutrino sector of the theory.  An important point is that the hierarchy 
of entries in Eqs.~(\ref{eq:texture1}) and (\ref{eq:texture2}) can be 
understood in terms of a sequential breaking of the original symmetry 
group $G_f$, through a series of subgroups at successively lower scales 
below $M_F$.   We may assume that each of these scales
is associated with one or more $\Phi$ fields that acquire the most 
general set of vevs consistent with with the unbroken flavor symmetries 
at that scale; elements of the Yukawa matrices that differ hierarchically 
originate from the vevs of different $\Phi$ fields.  The textures in 
Eqs.~(\ref{eq:texture1}) and (\ref{eq:texture2}) then arise naturally given 
the symmetry breaking pattern shown in Table.~1.   
\begin{table}
\begin{center}
\begin{tabular}{cc} 
& U(3)$_Q\times$U(3)$_U\times$SO(3)$_D$ \\ \hline \hline
$h_t$ &  U(2)$_Q\times$U(2)$_U \times$U(1)$_{Q_3-U_3}\times$SO(3)$_D$ \\
$h_b$ &  U(2)$_Q\times$U(2)$_U\times Z_2$ \\
$h_c$ &  U(1)$_{Q_1} \times $U(1)$_{U_1} \times $U(1)$_{Q_2-U_2}\times Z_2$ \\
$h_s$, $h_b V_{cb}$ & U(1)$_{Q_1}\times$U(1)$_{U_1} \times Z_2$ \\
$h_s\lambda$, $h_b V_{ub}$ & U(1)$_{U_1} \times Z_2$ \\
$h_d$ & U(1)$_{U_1}$ \\
$h_u$ & nothing \\ \hline 
\end{tabular}
\end{center}
\caption{Sequential symmetry breaking leading to the mass matrices in 
Eqs.~(\ref{eq:texture1}) and (\ref{eq:texture2}) The U(1)$_{X_i}$ act on
the $X$ superfield of the $i^{th}$ generation, and the $Z_2$ flips the
sign of the superfield $D_1$.}
\end{table}

We may now treat the Yukawa couplings in Eqs.~(\ref{eq:texture1})
and (\ref{eq:texture2}) as small symmetry breaking parameters, and 
estimate the size of $G_f^{eff}$-violation in any $1/M_F$ suppressed operators
of interest in the low-energy effective theory.  This will allow us to
determine the sizes of squark, slepton, and neutrino nondegeneracies. 
The nonrenormalizable operators that we study may arise in a theory that 
is renormalizable at high energies if we work below the scale at which
heavy states have been integrated out, as in the approach of Froggatt 
and Nielsen \cite{fn}. In our model, we may implement this mechanism by 
introducing three 15-plet generations of vector-like fields at 
the scale $M_F$ that we integrate out at lower scales.  Denoting these 
heavy fields with the superscript $H$, the desired interactions are 
obtained providing that $D^H$, $U^H$, and $L^H$ transform as fundamentals 
under U(3)$_Q$, while $E^H$ and $Q^H$ transform under SO(3)$_D$.  For 
example, for the leptons, we have superpotential interactions of the form
\begin{equation}
W= \overline{L}^H \Phi_d L + L^H H_d E + \overline{E}^H \Phi_d E + E^H H_d L 
+M_F (\overline{E}^H E^H + \overline{L}^H L^H) \,\,\, ,
\end{equation}
which produce the charged lepton Yukawa matrix via the two superfield diagrams
shown in Figure~1.  The fact that the Yukawa interactions arise via
renormalizable interactions will affect the form of the
neutrino mass matrices, as we shall see shortly.

\begin{figure}[t]
\centerline{\psfig{file=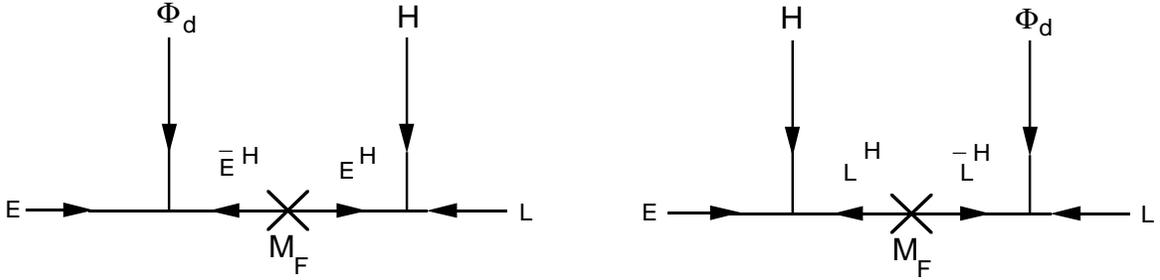,width=\textwidth}}
\caption{Superfield diagrams contributing to the charged lepton Yukawa
matrix.}
\end{figure}

We now consider the phenomenology of the neutrino sector of the theory.
The neutrino superpotential is given by
\begin{equation}
W = \rho \, \nu \nu + L H_u \nu \,\,\, ,
\label{eq:nusup}
\end{equation}
where $\rho$ is some field whose (B-L breaking) vev determines the
right-handed neutrino mass scale.  We have explicitly constructed our model so
that these interactions are proportional to the identity in flavor space, at
lowest order.  To find a correction to this result, we must be able to
construct a symmetry-breaking operator out of the $\Phi$ that transforms
nontrivially under SO(3)$_D$  alone, and is in the product ${\bf 3} \times
{\bf 3}$.  Because of the transformation properties of $\Phi_d$ under
U(3)$_Q$, the only way we can form a U(3)$_Q$ singlet is through the product
$\Phi_d^\dagger \Phi_d$.  However, while an operator of the form
\begin{equation}
\frac{1}{M_F^2} L H_u \Phi_d^\dagger \Phi_d \nu
\label{eq:not}
\end{equation}
is invariant under the flavor symmetry, it is not a holomorphic function of 
the fields, and cannot be included in the superpotential.  Thus, there is no 
way we can directly modify Eq.~(\ref{eq:nusup}) by higher-dimensional, 
$\Phi$-dependent terms in a way that is consistent both with $G_f^{eff}$ 
invariance, and unbroken supersymmetry.

\begin{figure}[t]
\centerline{\psfig{file=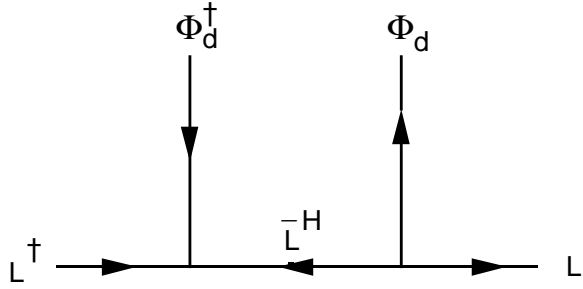,width=8cm}}
\caption{Superfield diagram contributing to Eq.~(\ref{eq:correc}).}
\end{figure}

However, there is an another way in which the exact neutrino degeneracy in 
Eq.~(\ref{eq:nusup}) may be perturbed, namely through corrections to the
K\"ahler potential of the theory, which is not a holomorphic function of the
fields.  Let us represent the U(2)$_Q$ singlet and doublet parts of $\Phi_d$ 
as $\Phi_d^{(1)}$ and $\Phi_d^{(2)}$, respectively.  Corrections to the 
minimal K\"ahler potential for the left-handed neutrinos of the form
\begin{equation}
K = L^\dagger L + \frac{1}{M_F^2} L^\dagger \Phi^{(1)\dagger}_d \Phi^{(1)}_d L
+  \frac{1}{M_F^2} L^\dagger \Phi^{(2)\dagger}_d \Phi^{(2)}_d L  \,\,\, ,
\label{eq:correc}
\end{equation}
force a field redefinition to place the neutrino kinetic terms in canonical
form.  The correction terms in Eq.~(\ref{eq:correc}) may arise in our 
Froggatt-Nielsen model via the superfield diagram in Figure~2.  If we write 
the lowest order field redefinition as $L \rightarrow (1-\Delta/2)L$,  with
the matrix $\Delta$ determined from Eq.~(\ref{eq:correc}), then the
original neutrino mass matrix $M_{LL}$, which is proportional to the 
identity, becomes  $M_{LL} [1-(\Delta-\Delta^T)/2]$.  Using the 
$\{ \Phi \}$ vevs given in Eq.~(\ref{eq:texture1}), we may now
construct the neutrino mass matrix.  We find
\begin{equation}
M_E  \sim  \Phi_d^{(1)\dagger}+ \Phi_d^{(2)\dagger} \approx
\left(\begin{array}{ccc} 
h_d & h_d & h_d \\
h_s \lambda & h_s & h_b \\
h_b V_{ub} & h_b V_{cb} & h_b \end{array}\right)
\label{eq:me}
\end{equation}
\[
M_{LL}  \sim {\bf 1} +
{\Phi_d^{(1)\dagger}}' {{\Phi_d}^{(1)}}' + {\Phi_d^{(2)\dagger}}' 
{\Phi_d^{(2)}}' 
\]\begin{equation}\approx
{\bf 1}+\left(\begin{array}{ccc} 
h^2_d & a h_d h_b + h_d h_s & b h_d h_b  + h_d h_b V_{cb} \\
a h_d h_b + h_d h_s & a h_b^2 + h_s^2  & ab h^2_b \\
b h_d h_b + h_d h_b V_{cb}  & ab h^2_b & b^2 h_b^2 \end{array}\right) \, .
\label{eq:mvll}
\end{equation}
Here we have included the order one coefficients $a$ and $b$ where they
are relevant to our results.  We consider these matrices in detail
below.

\section{Neutrino Phenomenology}

{\em Neutrino mass splittings.} Clearly the neutrino mass splittings are 
controlled by the correction to the identity matrix shown in 
Eq.~(\ref{eq:mvll}).  Notice that the leading, ${\cal O}(h_b)$, correction 
arises from the coupling of a single linear combination of the 
$\Phi_d^{(1)}$ fields, ${\Phi_d^{(1)}}' \sim (h_d, h_b, h_b)$, that 
participates in the $\overline{L}^H \Phi L$ vertex in Figure~2.   As a 
consequence, the 2-3 block in Eq.~(\ref{eq:mvll}) must have one vanishing 
eigenvalue, which is corrected at order $h_s^2$.  The extent to which this 
eigenvalue relation can be preserved will be discussed in the section on 
fine-tuning below.  The ratio of mass splitting is given by
\begin{equation}
\frac{\Delta m^2_{12}}{\Delta m^2_{23}} = \frac{(1+h_s^2)^2 - (1+h_d^2)^2}
{(1+h_b^2)^2 - (1+h_s^2)^2} \approx h_s^2/h_b^2 \sim 0.002 \,\,\, .
\label{eq:m2ratio}
\end{equation} 
This is roughly consistent with the ratio preferred by the solar and
atmospheric neutrino problems \cite{superk,bk},  
$\Delta m^2_{12}/\Delta m^2_{23} 
= 0.8 \times 10^{-5} \mbox{eV}^2 /0.5 \times 10^{-2} \mbox{eV}^2 \sim 0.0016$,
where we assume the small angle MSW solution to the solar neutrino problem.  
The overall mass scale can be set by an appropriate  choice for the $\rho$ 
vev, which is a free parameter in the theory. 

{\em Neutrino mixing angles.}  To obtain nonvanishing neutrino mixing angles,
we require nonzero relative rotations in diagonalizing $M_E$ and $M_{LL}$.  
Although the matrices in Eqs.~(\ref{eq:me}) and (\ref{eq:mvll}) are built
out the same set of primordial symmetry-breaking fields, we avoid any unwanted
alignments by assuming that there are generally more than one field involved
in symmetry breaking at any given scale; entries of comparable size in
$M_{E}$ and $M_{LL}$ are then expected to differ by order one factors.
For example, consider the leading corrections, of order $h_b$, originating
from a number of fields transforming like $\Phi_d^{(1)}$.  While one linear 
combination of these fields, ${\Phi_d^{(1)}}'$, is involved in $M_{LL}$ as 
we saw above,  $M_E$ originates from the sum of two different superfield 
diagrams, shown in Figure~1.  The second diagram involves different vertices, 
and thus does not depend on the same linear combination of the 
$\Phi_d^{(1)}$.  Thus, it is safe to read off the order of magnitude of the 
neutrino mixing angles from these matrices.  All terms proportional to
$h_b$ in Eq.~(\ref{eq:mvll}) (including those in the 12 and 13 entries) 
are diagonalized by an order one 23 rotation, $\sin^2 2\theta_{23} \sim 1$.
The 12 block of the resulting matrix is then diagonalized with
$\sin^2 2\theta_{12} \sim 4 h^2_d/h^2_s\sim 0.009$.  The small 
angle MSW solution to the solar neutrino problem requires  
$\sin^2\theta_{12}= 3\times 10^{-3}$--$1.1\times 10^{-2}$ \cite{bk}, while the 
range for $\sin^2 2\theta_{23}$ preferred by the atmospheric neutrino anomaly 
is $\sin^2 2\theta_{23}=0.4$--$0.6$ \cite{nra}.  Thus, the model gives the 
desired order of magnitude relations.  

{\em Fine-tuning.}  We have seen that our results for neutrino mass squared 
splittings and mixing angles followed from the observation that the 
largest eigenvalues of the nontrivial matrix in Eq.~(\ref{eq:mvll}) were of 
order $h_b^2$ and $h_s^2$, rather than both $h_b^2$.  This result was 
a consequence of the renormalizable origin of the flavor symmetry breaking
operators.  While our result does not appear to involve any fine-tuning,
we will now consider more precisely under what circumstances this is 
actually the case.

First, there was an assumption implicit in the analysis, that the 23 and 33 
elements of $\Phi$ have no relative phase.  If this were not the case, we 
would find that the order $h_b^2$ correction to $M_{LL}$ in $\Delta+\Delta^T$ 
would again have two eigenvalues of order $h_b^2$.  Since the origin of CP 
violation is unclear, we could simply resign ourselves to fact that the 
absence of this relative phase is a restriction that we must place on the 
model.   However, in models where CP is a good symmetry at high energies, 
and then spontaneously broken at some scale $\Lambda_{CP}$, it may simply 
be the case that $\Lambda_{CP}$ lies below the scale at which $h_b$ is 
generated.  Then phases may arise elsewhere in the Yukawa textures, but will 
never spoil our results.   The second, and 
somewhat more subtle point, is that the eigenvalue result follows because 
we have a nongeneric set of operators involving the $\Phi_d^{(1)}$ fields 
immediately below the scale where we integrate out the vector-like states.  
Since the K\"ahler potential is not protected by the supersymmetric 
nonrenormalization theorem, one might worry that we generate operators 
involving other combinations of the $\Phi_d^{(1)}$ fields when we run down 
to the scale $h_b M_F$ where these fields acquire vevs, and are themselves 
integrated out of the theory.  There are a number of ways to address 
this point.  First, it may be the case that $h_b$ is actually generated at the 
scale $M_F$ rather than below it;  models exist in the literature in which 
$\langle \Phi \rangle$ is generated at the scale $M_F$, but of order 
$M_F/16\pi^2$, as a consequence of finite loop effects.   The remaining 
Yukawa couplings may then follow from a sequential breaking of the 
remaining flavor symmetries at lower scales, in the way suggested earlier.  
Another response is to point out that there are no fields left in our model 
below the scale $M_F$ that can couple to $\Phi_d$ and contribute to 
wavefunction renormalization that may generate unwanted operators; this may 
persist an extensions of the theory that explains the origin of 
the $\Phi$ vevs, though extending the model in this way is well 
beyond the scope of this letter.  Whether the reader chooses to view the 
neutrino mass relations as a prediction of the theory, as we have suggested, 
or take them as inputs to which undetermined coefficients are fit in a 
phenomenological approach, the flavor structure of the neutrino sector in 
this scenario leads to interesting predictions for the flavor structure of 
the squark and slepton mass matrices, as we will now discuss.

\section{The quark and lepton sector}  
Like many other models with 
non-Abelian family symmetries, this model naturally solves the supersymmetric 
flavor problem.  The values of the $K$-$\overline{K}$ mass difference and of 
$\epsilon^{\prime}/\epsilon$ imply that the masses of the down and strange 
squarks are very nearly degenerate if the quark and squark mass matrices 
are not aligned.  In this model, the squark masses of the first two 
generations are degenerate in the flavor symmetric limit, and are 
split only by effects of the small fermion Yukawa couplings.   All
flavor-changing neutral current processes in the quark and charged lepton
sectors are calculable in terms of the entries in Eqs.~(\ref{eq:texture1}) 
and (\ref{eq:texture2}), which were determined in part by the requirement 
we obtain the neutrino phenomenology described in the previous section.

A comprehensive analysis of the flavor changing neutral current constraints
on soft masses in a general supersymmetric theory has been given by 
Gabbiani, et al.\cite{gabbiani}.   The constraints are presented in terms
of the quantities $(\delta_{ij})_{AB}$, the ratio of the $ij$ off-diagonal 
term in a given squark or slepton mass-squared matrix to the average
diagonal entry, in the basis where the fermion mass matrices are diagonal. 
(Since the scalars are nearly degenerate, the ambiguity in the 
word ``average" is irrelevant).  The indices $A$ and $B$, are either $L$ 
or $R$, indicating the helicity of the corresponding quark or lepton. 
The order of magnitudes for the $\delta_{ij}$ can be determined by 
constructing all the $1/M_F$ suppressed operators that contribute to the 
soft scalar masses, and to the trilinear interactions involving the Higgs 
fields (the A-terms).  For example, we find the scalar mass squared matrices
\begin{equation}
m^2_{\tilde{Q}} \sim \left(\begin{array}{ccc}
m_1^2  + h_s^2 \lambda^2 m^2 & h_c h_s \lambda m^2 & h_s \lambda m^2 \\
h_c h_s \lambda m^2 & m_1^2 + h_c^2 m^2 & h_s m^2 \\
h_s \lambda m^2 &  h_s m^2 & m_3^2 \end{array}\right)  
\end{equation}
\begin{equation}
m^2_{\tilde{U}} \sim 
\left(\begin{array}{ccc}
m_1^2  + h_u^2 m^2 & h_u h_c  m^2 & h_u m^2 \\
h_u h_c m^2 & m_1^2 + h_c^2 m^2 & h_s m^2 \\
h_u m^2 &  h_s m^2 & m_3^2 \end{array}\right) \, ,
\end{equation}
where the $m$ and $m_i$ are generic supersymmetry-breaking masses.
The mass matrix for $m^2_{\tilde{D}}$ is of the same form as 
Eq.(\ref{eq:mvll}),while $m^2_{\tilde{E}}$ and  $m^2_{\tilde{L}}$ have the 
same form as $m^2_{\tilde{Q}}$ and  $m^2_{\tilde{D}}$, respectively.  Thus 
in our model, the severely constrained combination, 
$[(\delta^d_{12})_{LL}(\delta^d_{12})_{RR}]^{1/2} 
\sim (\lambda h_c h_s \cdot h_d h_s)^{1/2} \sim 10^{-7}$, while the
bound from $K$-$\overline{K}$ mixing is of order $10^{-3}$ \cite{gabbiani}.
We find that the remaining $\delta_{LL}$ and $\delta_{RR}$ are so 
much smaller than the constraints in Ref.~\cite{gabbiani}, that we 
won't bother stating them explicitly.  On the other hand,  we find that the 
constraints from the $(\delta_{ij})_{LR}$ are much more significant.  The 
left-right scalar mass terms arise from the trilinear scalar interactions 
involving the Higgs fields.   Since these interactions have the same 
$G_f^{eff}$ symmetry structure as the Yukawa couplings themselves, the 
corresponding scalar mass squared matrices have the same structures as 
Eqs.~(\ref{eq:texture1}) and (\ref{eq:texture2}), with an overall 
scale $A\langle H\rangle$, where $A$ has dimensions of GeV.  The most 
significant constraints on the model 
from the $(\delta_{ij})_{LR}$ are given in Table~2, assuming $A$-parameters 
of 100 GeV.  We translate the constraints on the $\delta_{ij}$ into lower 
bounds on the average quark or slepton mass. In the slepton sector, the 
strongest bound comes from $\mu\rightarrow e\gamma$, which forces the 
slepton mass to be greater than approximately $400$ GeV.  Since we expect 
all supersymmetric particle masses to be less than O(1) TeV, this would 
imply that an eventual improvement in the bound would rule out this model.  
Note however that the bound can be further relaxed if we take into
account possible variation in the unknown order one coefficients that 
multiply the $1/M_F$ suppressed operators.  Such fluctuations are in fact 
necessary elsewhere in the model, to reproduce the Georgi-Jarlskog factors 
of $3$ in the charged lepton Yukawa matrix.  This limitation is also found 
in other models of flavor that cannot be embedded into a grand unified 
theory (e.g. Ref.~\cite{chm1})\footnote{Thus, these 
models are consistent with a string unification of the gauge couplings, 
rather than a field-theoretic unification.}.   If $\mu\rightarrow e \gamma$ 
is somewhat suppressed relative to our naive estimates, then the next 
signature of new physics would be $\tau\rightarrow\mu\gamma$, which should 
be seen with an order-of-magnitude improvement in experimental sensitivity.  
Note that potentially stronger bounds on the squark masses can be derived from 
$\epsilon'/\epsilon$, but only if assumptions are made about 
unknown CP-violating phases.  In the case in which CP violating phases
are of order one, we find, for example, that 
${\rm Im} (\delta^d_{12})_{LR} \sim (A \langle H \rangle / m^2) h_s \lambda
\sim [(200 \mbox{GeV})(200 \mbox{GeV})/(500 \mbox{GeV})^2 ]h_s \lambda 
\sim 2 \times 10^{-5}$, which is in borderline agreement with the
experimental bounds for $500$ GeV gluinos.  

\begin{table}
\begin{center}
\begin{tabular}{llll}
\multicolumn{1}{c}{Process}& \multicolumn{1}{c}{Expt. Constraint} & 
\multicolumn{1}{c}{Model Prediction} & \multicolumn{1}{c}{Bound}\\ \hline
$\mu\rightarrow e\gamma $&$(\delta^{\ell}_{21})_{LR}< (1.4-3.8)\times 10^{-6}
$&$ (\delta^{\ell}_{21})_{LR}={\lambda m_s A \over m^2_{\tilde{\ell}}}$&$
m_{\tilde{\ell}}> (330-420)\ {\rm GeV}$\\
$\tau\rightarrow \mu\gamma$&$ (\delta^{\ell}_{23})_{LR}< 
(1.7-4.4)\times 10^{-2}$ & $
(\delta^{\ell}_{23})_{LR}={m_b A \over m^2_{\tilde{\ell}}}$&$
m_{\tilde{\ell}}> (100-130)\ {\rm GeV}$\\
$b\rightarrow s\gamma$&$ (\delta^q_{32})_{LR}< (1.3-3)\times 10^{-2} $ & $
(\delta^q_{32})_{LR}={m_b A \over m^2_{\tilde{q}}}$&$
m_{\tilde{q}}> (250-300)\ {\rm GeV}$\\
\end{tabular}
\caption{Most significant $(\delta_{ij})_{LR}$.  The experimental
constraints shown are given for $m_{\tilde{l}}=100$ GeV 
and $m_{\tilde{q}}=500$ GeV, and scale as the square of the scalar mass.
The ranges in values shown corresponds to $m_{\tilde{g}}/m_{\tilde{q}}$ and 
$m_{\tilde{\gamma}}/m_{\tilde{\ell}}$ varying between 0.3 and 5.  The
bounds correspond to $A=100$ GeV.}
\end{center}
\end{table}
In addition to the predictions for flavor-changing neutral currents, this model
makes a striking prediction for the diagonal squark and slepton
masses.  Since the low-energy effective theory is constrained by the symmetry 
$G_f^{eff}$, whose non-Abelian factors are 
U(2)$_Q \times$U(2)$_U \times $SO(3)$_D$, we expect the left-handed up and 
down squarks, as well as the right-handed up squarks to be nearly degenerate 
for the first two generations respectively (with the third generation scalar 
mass differing by some order one factor), while the right-handed down 
squarks remain more nearly degenerate for all three generations, because of 
the approximate SO(3) symmetry.  Corresponding statements can be made in the 
lepton sector.  If we choose to think of the high energy flavor symmetry 
as $G_f$, then we can restate this result by noting that the original 
symmetry, which treats all three generations equally, has some remnant 
effect in sectors of the theory that are not directly affected by the 
mechanism which generates the top quark Yukawa coupling. This fact is 
what allowed us to obtain quasi-degenerate neutrinos, as well as the 
associated phenomenology described above.

\section{Conclusions}
We have presented a model with quasi-degenerate neutrinos that can
account for the solar and atmospheric neutrino problems.  By working with
a flavor symmetry that constrains both the quark and lepton sectors,
the model relates the neutrino mass squared splittings and mixing angles to
quark Yukawa couplings, $\Delta m^2_{12} / \Delta m^2_{23} \sim m_s^2/m_b^2$, 
$\theta_{12} \sim m_d/m_s$ and  $\theta_{23} \sim 1$.  These have the
correct orders of magnitude to give us the small angle MSW solution to
the solar neutrino problem, as well as atmospheric $\nu_\mu$-$\nu_\tau$
oscillations.   The model naturally evades the supersymmetric flavor problem by
maintaining sufficient squark and slepton degeneracy,  while yielding definite 
predictions for the flavor structure of the superparticle mass matrices.  In 
particular, the approximate flavor symmetry at low energies implies 
approximate three-generation degeneracy for the right-handed down squarks and 
left-handed charged leptons, respectively, while approximate degeneracy 
among only the first two generations for the remaining squarks and slepton 
states.  The model predicts $\mu\rightarrow e \gamma$ just beyond the 
current bound, so that we would expect this process to
be detected with an order of magnitude improvement in experimental 
sensitivity.



\end{document}